# Crystal orientation and grain size: do they determine optoelectronic properties of MAPbI$_3$ perovskite?

Loreta A. Muscarella[1], Eline M. Hutter[1], Sandy Sanchez[2], Christian D. Dieleman[1], Tom J. Savenije[3], Anders Hagfeldt[2], Michael Saliba[4], Bruno Ehrler[1,*]

[1] Center for Nanophotonics, AMOLF, Science Park 104, 1098 XG Amsterdam, The Netherlands

[2] Laboratory of Photomolecular Science (LSPM) École Polytechnique Fédérale de Lausanne (EPFL), Station 6, 1015 Lausanne, Switzerland

[3] Department of Chemical Engineering, Delft University of Technology, Van der Maasweg 9, 2629 HZ Delft, The Netherland

[4] Institute of Materials Science Technical University of Darmstadt Alarich-Weiss-Strasse 2, D-64287 Darmstadt, Germany

**Abstract**

It is thought that growing large, oriented grains of perovskite can lead to more efficient devices. We study MAPbI$_3$ films fabricated via Flash Infrared Annealing (FIRA) consisting of highly oriented, large grains. Domains observed in the SEM are often misidentified with crystallographic grains, but SEM images don't provide diffraction information. We measure the grain size, crystal structure and grain orientation using Electron Back-Scattered Diffraction (EBSD) and we study how these affect the optoelectronic properties as characterized by local photoluminescence (PL) and time-resolved microwave conductivity measurements (TRMC). We find a spherulitic growth yielding large (tens of µm), highly oriented grains along the (112) and (400) planes in contrast to randomly oriented, smaller (400 nm) grains observed in films fabricated via conventional antisolvent (AS) dripping. We observe a local enhancement and shift of the photoluminescence emission at different regions of the FIRA clusters, but these can be explained with a combination of light-outcoupling and self-absorption. We observe no effect of crystal orientation on the optoelectronic properties. Additionally, despite a substantial difference in grain size between our FIRA sample and a conventional AS sample, we find similar photoluminescence and charge carrier mobilities and lifetime for the two films. These findings show that the optoelectronic quality is not necessarily related to the orientation and size of crystalline domains in perovskite films indicating that fabrication requirements may be more relaxed for perovskites.

**Introduction**

Hybrid perovskites have recently gathered significant attention due to the high efficiency of perovskite-based solar cells and other optoelectronic devices[1,2]. One of the most surprising properties of these materials is that the performance is very tolerant to different methods of fabrication[3,4,5], different compositions[6,7], and chemical treatments[8,9,10]. This is reflected in high photoluminescence quantum efficiencies (PLQE)[11,12], which is a measure of the fraction of radiative versus non-radiative decay, and therefore a direct measurement of the optical quality. In solar cells, the PLQE, for example, is directly related to the open-circuit voltage[13].

In practice, PLQE is reduced by the presence of defects[13,14] which are often related to chemical impurities such as interstitials, vacancies, dangling bonds[15,16] or defects on the surface and grain boundaries[17,18,19]. Both bulk and surface defects have been extensively studied in perovskites[20,21,22,23], and efficient passivation strategies are now routinely employed to achieve high LED and solar cell efficiencies[24,25,26]. In addition, chemical methods (e.g. Lewis bases[21] or chloride-based additives[27,6]) in the perovskite precursor are often applied to grow larger grains which has been thought to suppress non-radiative recombination pathways by reducing the number of grain boundaries[28,29,30]. Furthermore, these changes in the synthesis route affect the crystal growth and therefore the preferred crystallographic orientations[31,8,32].

However, it is unclear if the changes in grain size and orientation obtained by these treatments cause the improved optoelectronic properties, or if these are mainly related to passivation effects from the additives. Crystallographic orientation and their relation to the photoluminescence and other properties have not been studied in thin films so far because spatial resolution of the crystallographic parameters was lacking. The morphological "grain" observed in SEM images does not necessarily correspond to a crystallographic grain; so additional information is needed in order to relate the grain size with optoelectronic properties.

Here we use EBSD to measure size, orientation and rotation of crystallographic grains in polycrystalline MAPbI$_3$ films with high spatial resolution. We study a MAPbI$_3$ thin film where crystallization is induced by FIRA, a low cost and rapid synthesis method[33,34]. [35]These films exhibit highly oriented (112) and (400) planes with large grain size (tens of micrometers). We find that the growth is spherulitic, i.e. needle-like arrays, yielding ~100 micrometer sized clusters that consist of radially grown grains. With EBSD mapping we find that the two crystal orientations in the FIRA films are well-separated in pairs in the large clusters of grains. We compare the PL from these clusters and find that PL intensity and spectrum is the same for the two crystal orientations. These results suggest that the crystallographic orientation does not govern the optoelectronic quality of perovskite thin films. Furthermore, we find enhanced emission and a red-shift at the cluster boundaries and at the nucleation sites, which we attribute to favorable light-outcoupling and self-absorption. Finally, we compare the FIRA sample to one where crystallization is induced by the conventional AS dripping method, from the same precursor solution. This method produces sub-micron grains with random orientation. Both samples show comparable charge carrier mobility and lifetime demonstrating that these properties are not necessarily determined by the grain size, at least for grains above a few hundred nanometers.

**Results and Discussion**

To study the relation between the perovskite crystal orientation and its optoelectronic properties we first synthesize MAPbI$_3$ on ITO via FIRA wherein the spin coated perovskite film is annealed using a short (1.2 second), highly intense infrared illumination to induce nucleation, as previously reported[33]. Importantly, to decouple the effect of preferential orientation and the presence of additives on to the optoelectronic properties, we fabricate a highly oriented FIRA samples without the aid of additives. For comparison we also fabricate a sample by the AS method where chlorobenzene is rapidly poured the liquid precursor while spin coating. Both the FIRA and AS sample were fabricated from the same precursors under identical conditions, and thus the only difference is the crystallization process.

We initially characterize the morphology of both samples using scanning electron microscopy (SEM) as shown in **Figure 1a,b** and **Figure S1**. SEM images show a dramatic difference in cluster size from the antisolvent method (100 nm - 2 µm) compared to FIRA (~100 µm). Often, these clusters seen in SEM images are assigned to grains. In crystallography, the term "grain" is defined by a coherently diffracting domain of solid-state matter which has the same structure as a single crystal[36]. Therefore, from SEM images alone it is not possible to define the apparent domains as crystallographic grains because diffraction information is not measured. For this reason, we use "clusters" to describe the large perovskite domains shown in SEM images.

We analyze the bulk crystal structure of the two systems deposited on ITO using X-Ray Diffraction (XRD). The AS sample shows a tetragonal XRD pattern where peaks from (110), (112), (220), (310) planes arise from the background[37,38]. In contrast, the FIRA sample shows a strong preferential orientation along the (112) and (400) planes (**Figure 1c**). A cut-off of the primitive tetragonal cell and the planes which show the highest diffraction peaks are shown next to the diffraction patterns.

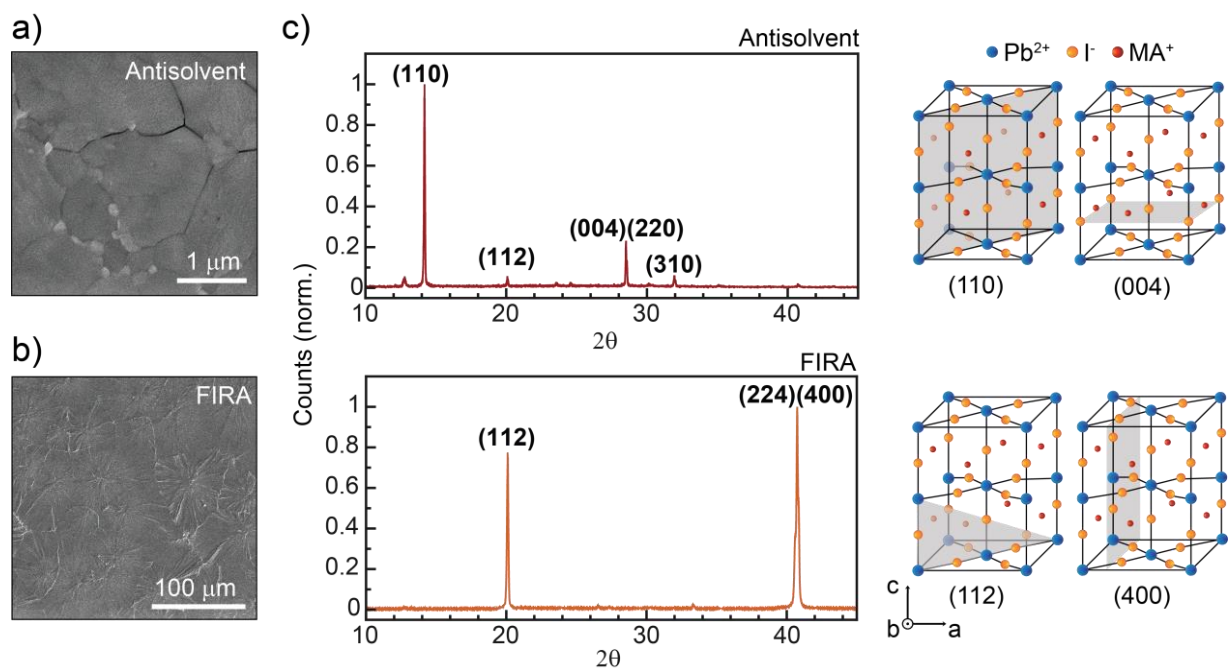

**Figure 1.** *Morphology and crystal orientation of AS and FIRA sample deposited on ITO.* SEM image of MAPbI$_3$ crystallized with **a)** AS and **b)** FIRA, **c)** XRD pattern of AS and FIRA MAPbI$_3$ with the corresponding unit cell cut along the (110) and (004) planes for the AS sample and along (112) and (400) for the FIRA sample.

To follow the crystal growth of the FIRA film, we apply infrared annealing with varying pulse duration (**Figure 2a**). After 0.2 s of annealing we observe a needle-like crystal morphology. After 0.5 s the crystals have grown in a space-filling manner by branching from the parental needle until each domain impinges with neighboring domains resulting in straight boundaries. Optical microscopy of the final films (**Figure 2b**) shows large perovskite domains in agreement with the SEM images. We use polarized light microscopy to find how the two orientations observed in the XRD are spatially distributed. **Figure 2b** shows the presence of paired oriented domains within FIRA films for all the clusters (see also **Figure S2**) originated by the change in polarization of the incident polarized light caused by the two different refractive indices in the different crystal directions. This is a indication of spherulitic growth (schematically shown in **Figure 2c)** via non-crystallographic branching[39] typical for many polymeric materials[40] and inorganic salts[41]. **Figure 2**Interestingly, there also seems to be a common direction to all clusters, suggesting a global effect from temperature or strain gradient. In general, spherulitic growth requires the use of a saturated solution, high viscosity, and slow crystallization. This growth is also catalyzed by the presence of impurities[39] or strain in the material. In this case, MAPbI3 and the ITO (or quartz) show a substantial difference in the thermal expansion coefficient ($\alpha_{MAPbI3}$ =6.1 × 10$^{-5}$ K$^{-1}$, $\alpha_{ITO}$ = 0.85 × 10$^{-5}$ K$^{-1}$, $\alpha_{glass}$ = 0.37 × 10$^{-5}$ K$^{-1}$ [42]), which has been shown to be the origin of strain during the cooling process after the thermal annealing[43]. Thus, strain can be considered as a factor inducing spherulitic growth in our system.

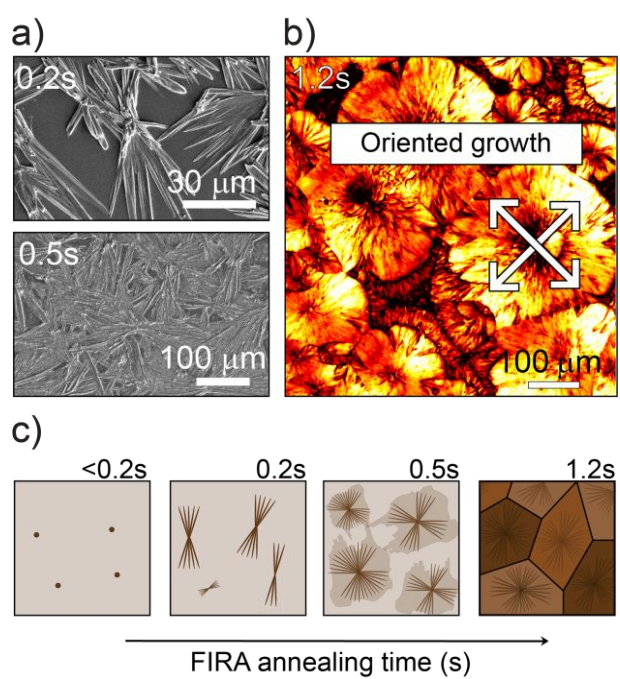

**Figure 2**. *Spherulitic growth mechanism of MAPbI$_3$ results in paired oriented domains.* **a)** SEM images of FIRA film after 0.2s and 0.5s of infrared annealing, **b)** Polarized microscopy image of the final FIRA film showing paired oriented perovskite domains, **c)** Schematic spherulitic growth of perovskite films after 0.2, 0.5, 1, and 1.2 s of FIRA annealing

To study the crystal grains and their orientation with high spatial resolution (10 nm) we use EBSD which is commonly used for investigations of grains in metal alloys[44], strain[45], and the nature of grain boundaries[46]. EBSD is a SEM-based technique where the incident electron beam, with a suitable voltage and current, interacts with a crystalline material and electron backscattered patterns, also called Kikuchi patterns, are produced by incoherent wide-angle electron scattering (thermal diffuse scattering) from a specimen. A scheme of the setup is depicted in **Figure 3a**. For a full description of EBSD measurements see **Supplementary Information S1**.

The main limitation which has restricted its application in the perovskite field is related to the high electron current needed for the phosphor screen to collect a sufficiently large signal-to-noise ratio. In case of perovskites containing organic cations, the use of a current of a few nA can already be damaging to the material. Recently, Adhyaksa et al.[47] have pioneered the application of EBSD for $MAPbBr_3$ using a direct electron detector which allows for low accelerating voltage of 5 kV instead of 30 kV, and low sample currents of pA instead of nA in conventional systems. We use the same detection system to collect the Kikuchi patterns from our $MAPbI_3$ films. The obtained Kikuchi patterns allow for the identification of grains, their size and shape and the nature of boundaries between them. By fitting the patterns, we can identify crystal phase, grain orientation and rotation, as described in **Supplementary Information S1**.

Importantly, since with EBSD diffraction information is measured, we can distinguish clusters from grains and define the crystallographic grain size of $MAPbI_3$ fabricated via FIRA and AS. **Figure 3b,c** shows an overlay of the image quality (brightness, IQ) with Inverse Pole Figure (color, IPF) of the AS and FIRA system along the z-axis. The IQ maps the sharpness of the Kikuchi lines obtained from the EBSD measurement which gives qualitative indications about the crystallinity of the material, topographic effects (e.g. roughness), strain of the microstructure and grain boundaries (for more details see **Supplementary Information S1**). The IPF represents the crystal orientation obtained from fitting the Kikuchi patterns to each pixel of the image with respect to a reference axis. Along the z-axis, the AS sample (**Figure 3b**), shows randomly oriented grains of hundreds of nanometers. On the contrary, in the FIRA sample (**Figure 3c**), all grains are aligned along [100] and [112] direction along the z-axis (green and purple color) and the two orientations are paired in larger regions, in agreement with XRD and polarized microscopy measurements. The distribution of orientations for the two samples clearly shows the mostly random orientation for the AS sample, and the bipolar distribution of orientations for the FIRA sample (**Figure 3d**). Consistent with XRD (**Figure 1c**), we observe significant orientation along both the [100] and [112] direction, but locally the ratio can vary (**Figure S3**). To investigate the actual grain size of the sample, we study the crystal orientation along the x- and y-direction for the FIRA sample (**Figure 3e,f** and **Figure S4** for orientation distribution). As the axes-dependent orientation is different for the three directions, we can now deduce that the 83% of the grains have an area between 0-30 $\mu m^2$ (see **Figure S5**) as shown in the EBSD, which does not coincide with the size clusters shown in SEM (**Figure 3g**). Furthermore, the orientation in the x- and y-direction is aligned along the growth of the initial needles, and the subsequent space-filling branches grow off these needles. The grain size obtained from EBSD for the AS sample is much smaller compared to the FIRA sample. with more than 90% of the grains smaller than 1 $\mu m^2$ (see **Figure S5**). Orientation maps along x- and y-direction for the AS sample are shown in **Figure S6**.

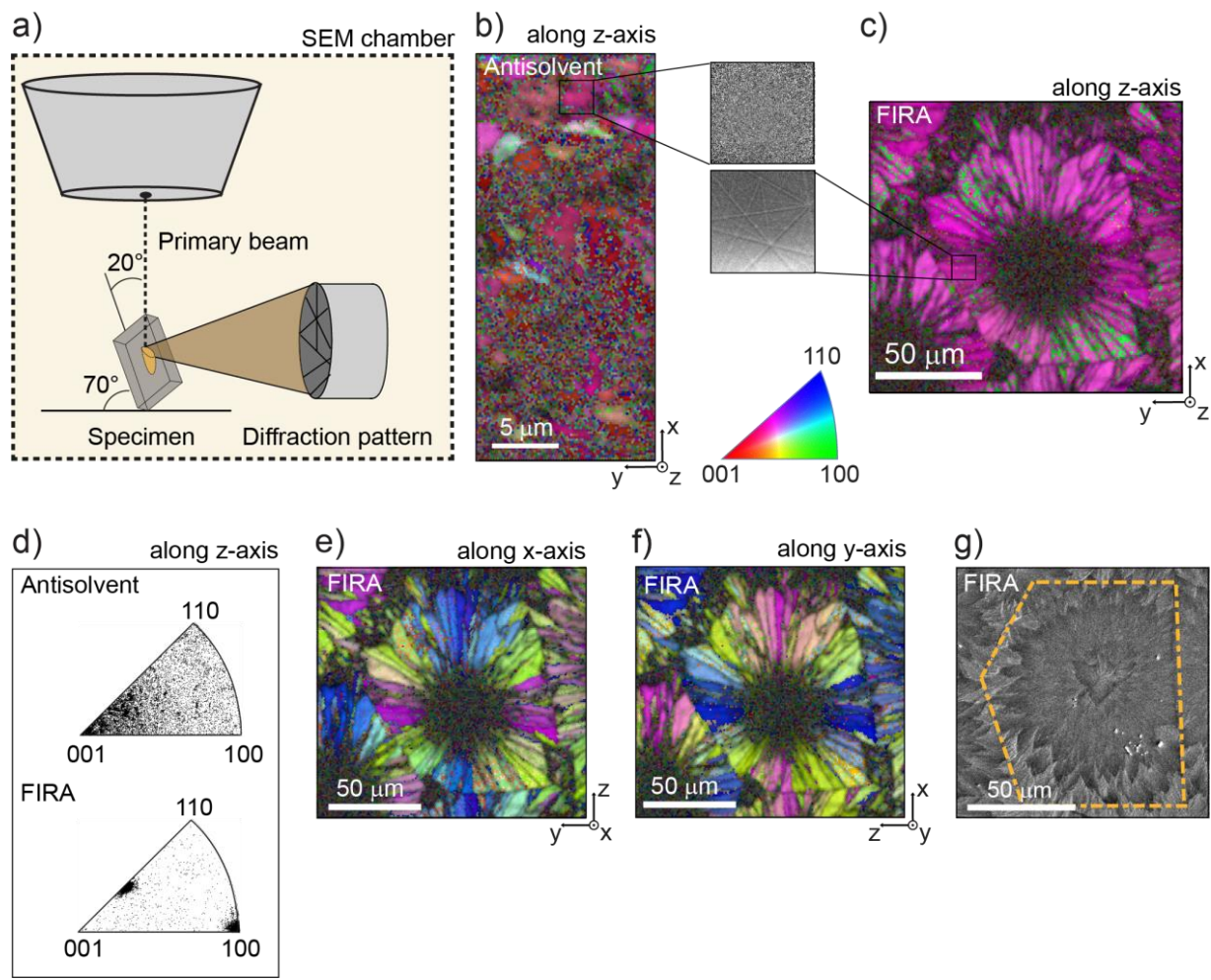

**Figure 3.** *EBSD maps reveal the crystal orientation and grain size of the two systems.* **a)** EBSD setup. **b)** Image Quality (IQ, brightness) overlay with Inverse Pole Figure (IPF) map of AS sample showing crystallographic orientation along z-direction. Inset, a magnification of the typical Kikuchi patterns recorded from the sample. **c)** Image Quality (IQ, brightness) overlay with Inverse Pole Figure (IPF) map of FIRA sample showing crystallographic orientation along z-direction. Inset, a magnification of the typical Kikuchi patterns recorded from the sample. **d)** Distribution of orientation for the two samples along z-direction. **e)** Image Quality (IQ, brightness) overlay with Inverse Pole Figure (IPF) map of FIRA sample showing crystallographic orientation along x-direction and **f)** y-direction. **g)** SEM image showing the apparent grain size of a FIRA cluster. The cluster measured with EBSD is highlighted with a dashed line.

MAPbI$_3$ has an anisotropic, tetragonal crystal structure, properties like trap-state density have been shown to be dependent upon crystal orientation[48,49]. Here, we study the optoelectronic properties of the two well-characterized and spatially separated orientations shown by FIRA sample using spatially resolved PL. We measure the PL intensity using a confocal imaging microscope using 405nm laser as excitation source with a power density of 0.23 W/cm$^2$. As the two orientations on the sample are spatially well-separated, we can map any difference in PL emission between them. We measure a large area of the FIRA and AS samples including a whole FIRA cluster (**Figure 4a** and **Figure S7** for a larger area). In the AS sample the PL intensity is relatively homogeneously distributed across the measured region, varying from cluster to cluster, consistent with many other works[50,51]. In contrast, the PL map of the FIRA sample shows an enhancement in intensity of two to six times at the cluster boundaries and at the nucleation site where the spherulitic growth is initiated. Enhancement in PL is often attributed to the presence of less non-radiative recombination, but light outcoupling must be taken in account as well. From AFM measurements, the AS sample shows only minor height variation between the center and the rest of the grain (RMS roughness 79 nm, **Figure 4b**). On the contrary, the FIRA sample shows significant height variation at the cluster boundaries (CBs), where boundaries from different clusters impinge with each other, and at the nucleation point (**Figure 4b** and in **Figure S8**). In these regions the film is around 400 nm to 1000 nm thicker compared to the interior of the cluster leading to a much larger roughness (RMS roughness 131 nm across the whole grain, between 85-65 nm in the interior of the cluster and 145-200 nm at the cluster boundaries). We note that FIRA clusters also show some local height variation in the interior part although less pronounced compared to the CBs. Thus, the rough nature of the boundary can favor light outcoupling, as shown on patterned perovskite surfaces[52]. We corroborate the assignment that the PL efficiency is constant across the FIRA clusters by measuring PL lifetime maps. The lifetime is constant across the cluster and CBs (with the exception of a few local hotspots) and hence there is no difference in radiative versus non-radiative rates at the darker and brighter regions (**Figure S9**). The PL enhancement observed can hence fully be explained by better light outcoupling. Next to these differences, there is no trend across the cluster that would correspond to the two different grain orientations. Thus, there is no direct correlation between the crystal orientation and the PL intensity.

The map of the PL peak position of the AS sample (**Figure 4c**) shows identical emission peak position for every grain. The FIRA sample shows a variation in emission wavelengths at different locations. The PL peak position at the CBs and nucleation point is red-shifted compared to the interior of the cluster (**Figure 4c**). Normalizing the PL spectra extracted from the boundary region of the map, we also observe asymmetric shape of the peak for the FIRA sample (**Figure S10**). This shape, in combination with the red shift of the peak has been assigned to self-absorption process when light travels through the perovskite layers before being emitted.[53]. We calculate the spectra expected from the emitted light passing through different thicknesses of $MAPbI_3$. We observe the red-shift at the boundaries and nucleation points corresponds to the light that has been transmitted through the 400nm – 800nm excess material as compared to the cluster interior region before being emitted (horizontal scale bar in **Figure 4c,** see **Supplementary Information S2** for details); this is in good agreement with the observed thickness variation. Again, we see no discrepancy for the regions that correspond to the two different, well-defined crystal orientations. Hence, the variation in PL emission intensity and wavelength cannot be correlated to the crystallographic orientation.

In **Figure 4d** we plot PL spectra from five random regions of the two samples. Here we show that the interior region of the FIRA cluster shows a comparable PL intensity with the AS (**FIRA Point 2&3**). Importantly, this shows that the PL emission is not solely determined by the grain size (at least for grains >400 nm).

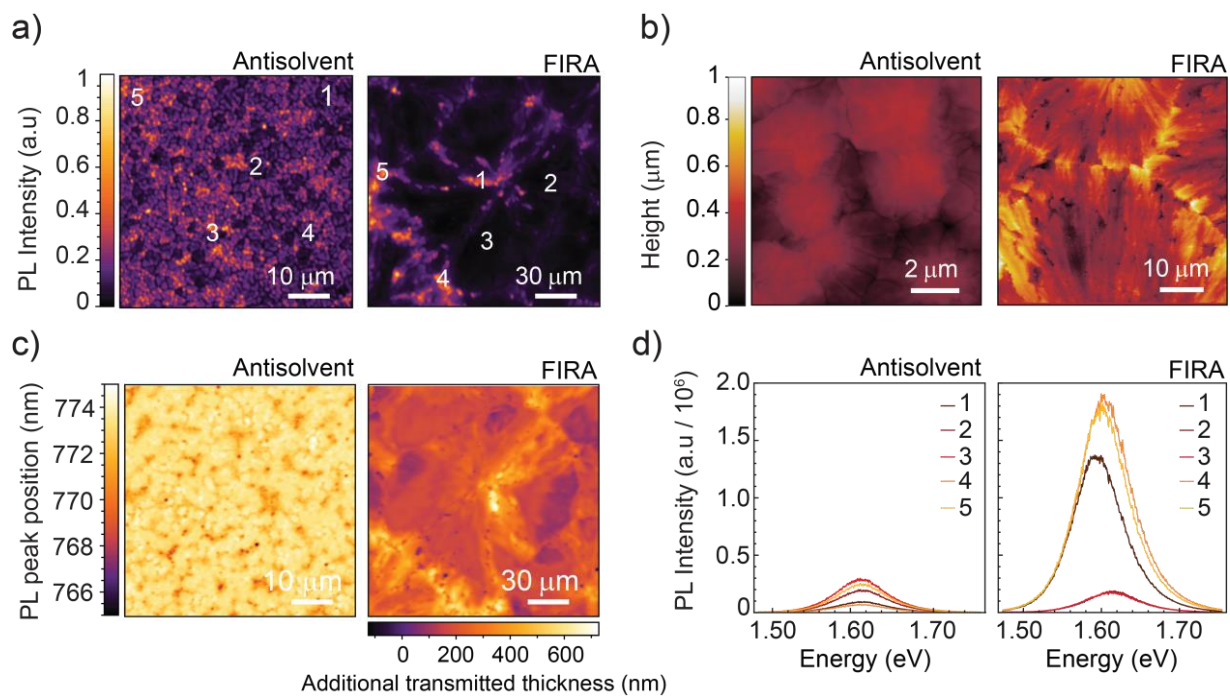

**Figure 4.** *Emission properties of MAPbI$_3$ varying crystal orientation and grain size.* **a)** Spatially resolved PL of AS and FIRA sample **b)** AFM image of AS and FIRA sample highlighting the cluster boundary region in the FIRA sample. **c)** Emission wavelength map of the same region as in figure (a) for the AS and FIRA sample. For the FIRA sample, the emission wavelength is converted into the additional thickness the light has been transmitted through, before it is emitted, **d)** PL spectra extracted from five random regions indicated in figure (a) in the AS sample showing similar PL intensity and no shift in the peak position, and PL spectra extracted from the cluster boundary and the inner cluster region of FIRA sample showing enhancement in PL at the cluster boundaries, and red-shift of the peak due to self-absorption.

Next to the optical properties, the electronic properties have been shown to depend on grain size in some cases[54]. To investigate the mobility and recombination dynamics of photoexcited charge carriers in our two systems we use the time-resolved microwave conductivity (TRMC) technique. The FIRA and AS samples were excited with 485 nm excitation wavelength. **Figure 5a,b** shows the photoconductance ΔG as a function of time after pulsed excitation of AS and FIRA sample, respectively. The product of the yield of free charges φ and their mobility Σμ (sum of electron and hole mobility) is derived from the maximum signal height ($\Delta G_{max}$) which was divided by the fraction of absorbed photons for the two samples to take in account difference in absorption. We find a mobility of (15±3) cm$^2$/(Vs) for the AS and (19±4) cm$^2$/(Vs) for the FIRA sample, which is comparable to sample-to-sample variation. The charge carrier lifetime is obtained from the photoconductance decay. The decay of the photoconductance represents the immobilization of free charges due to trapping or recombination. For both systems, we find that the lifetime of charges is in the order of a few hundred nanoseconds. We observe a slight increase of the effective mobility in the FIRA sample compared to the AS sample likely related to the enlarged grain size[54]. This difference is relatively small, despite the difference in grain size between the FIRA (tens of microns) and the AS sample (hundreds of nanometers). This shows that grain size does not play a major role in charge carrier transport properties. We note that the TRMC measurement mostly probes the local conductivity (~50 nm, more details about the probing length are reported in **Supplementary Note 3**). Inter-grain transport across larger distances may show larger differences in crystallographically different systems. This finding is consistent with the similar device performances[33] that have been reported for both FIRA and AS showing similar Jsc, Voc, FF and PCE. Practically, FIRA could allow a lower cost, environmentally friendly fabrication route to produce large scale and reproducible perovskite compared to the AS method[33].

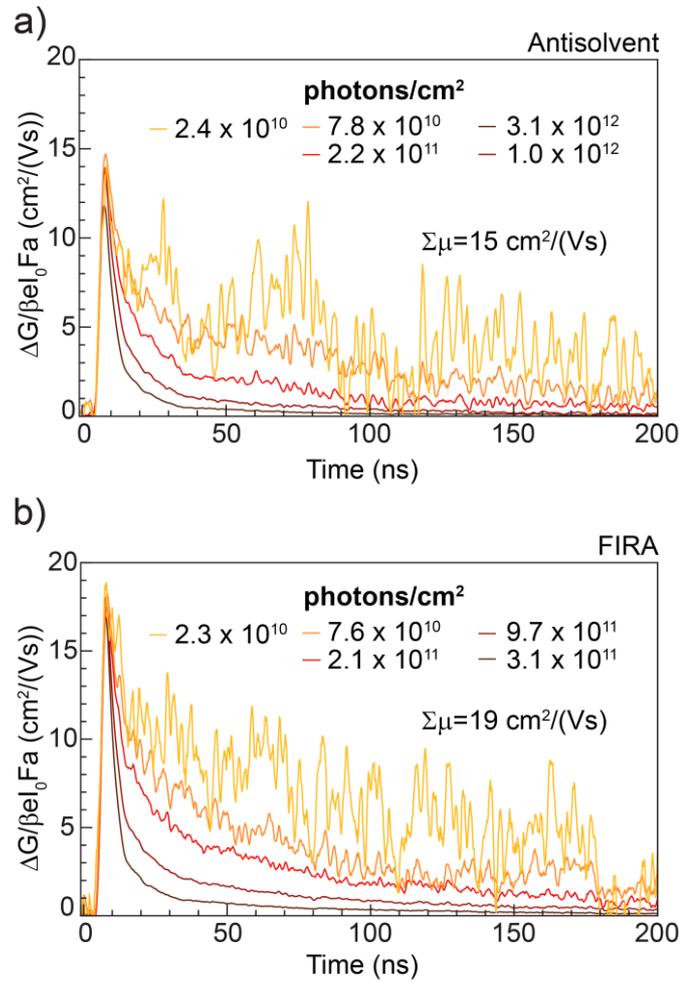

**Figure 5.** *Mobility and lifetime varying the grain size.* Time-resolved microwave conductivity (TRMC) traces measured at different carrier density for the **a)** AS and **b)** FIRA sample deposited on quartz.

## Conclusion

We have shown that the crystallographic orientation of MAPbI$_3$ grains does not determine the optical and local electronic properties. We study a MAPbI$_3$ thin film where crystallization is induced by FIRA. We apply EBSD to extract information about the microstructure of the perovskite thin film with high spatial resolution. The large FIRA clusters consist of grains that are tens of micrometer in size. They are highly oriented along the [112] and [100] direction perpendicular to the substrate. In comparison, a conventional sample fabricated via AS shows randomly oriented grains of hundreds of nanometers. We find increased PL intensity and a spectral shift in the FIRA sample compared to the AS sample, which we can be explained by roughness variations favoring light-outcoupling and self-absorption. Finally, we investigate charge carrier dynamics and find comparable lifetime and a slight increase in effective mobility in the FIRA and AS samples. We hence conclude that neither the grain size (when larger than a few hundreds of nm) nor the grain orientation are the dominant factor determining the optoelectronic properties of perovskite thin films. This finding implies that efforts towards a more efficient perovskite device may need to focus on reducing defects within the bulk and at the interface as well as impurities within the materials rather than growing large, oriented grains.

**Experimental methods**

**Thin film preparation**

The fabrication of the two systems (antisolvent dripping and FIRA method) is reported elsewhere[33].

**Thin Film Characterization**

The X-Ray diffraction pattern of perovskite films deposited on ITO was measured using an X-ray diffractometer, Bruker D2 Phaser, with Cu Kα 1.5406 Å as X-Ray source, 0.002° (2θ) as step size, 0.150 s as exposure time.

A FEI Verios 460 instrument was used to obtain SEM images. Atomic force microscopy (AFM) measurements were performed on a Veeco Dimension 3100 (Bruker) in tapping mode.

Optical microscope (Zeiss, AxioCam ICc 5) equipped with a 10x/0.2 objective EC Epiplan, polarizer and analyzer set at different angles was used for polarized optical microscopy image. We combine the optical microscope in reflection mode with two polarizers, one placed in the light path before the specimen, and a second one, called analyzer, between the objective lenses and eyepieces.

Steady-state photoluminescence (PL) of samples deposited on quartz was measured with a home-built setup equipped with a 640 nm continuous-wave laser as source of excitation (PicoQuant LDH-D-C-640) at a power output of 1 mW. Two Thorlabs filters, a long pass, ET655LP, and a notch, ZET642NF, were used to remove the excitation laser from the signal. The PL was coupled into a fiber connected to an OceanOptics USB4000 spectrometer. An integration time of 300 ms was used for each measurement.

For EBSD measurements, samples were deposited on ITO to avoid charging effects during the experiment. The detector used is a direct electron detector based on the Timepix sensor from Amsterdam Scientific Instruments (ASI). The best parameters for the scans were found to be 15 keV as voltage, 100 pA as current, 100 ms as exposure time and working distances between 12 mm and 10.1 mm. This translates to the application of 10 nAms electron dose per pixel which is around $10^3$ times lower compared conventional measurement reducing sample damage. The step size was chosen depending the cluster size shown by the specific sample, 200 nm and 1 μm for the AS and FIRA sample, respectively. EBSD data were collected using EDAX OIM software and a Python script was used for image processing. The resulting Kikuchi patterns were indexed using tetragonal symmetry, I4/mcm, using 1-3° as degree of tolerance. Detailed procedure for fitting the Hough's space is reported in the **Supplementary Note**.

Spatially resolved PL map was measured using a confocal imaging microscope (WITec alpha300 SR). A 405 nm laser diode (Thorlabs S1FC405) was used as excitation source where the PL intensity of the two $MAPbI_3$ films was collected in reflection mode through a NA 0.9 objective using a spectrometer (UHTC 300 VIS, WITec) leading to a spatial resolution of 0.33 µm. The intensity was measured within the 700 to 840 nm emission wavelength range. A 488 nm LP filter was used to remove the excitation laser from the signal. The light collection was done from the same sample side as the excitation. The PL spectra were converted to the energy scale using a Jacobian transformation[55].

Time-Resolved Microwave Conductivity (TRMC) was measured on AS and FIRA sample deposited on quartz. The thin films were placed in a sealed resonance cavity inside a nitrogen-filled glovebox to avoid degradation due to air exposure. The samples were excited at 485 nm using a pulsed excitation (10 Hz) and the photoconductance probed at different excitation density. Neutral density filters were used to vary the intensity of the incident light. The ΔG signal rise is limited by the width of the laser pulse (3.5 ns FWHM) and the response time of the microwave system (18 ns). The slow repetition rate of the laser of 10 Hz ensures full relaxation of all photo-induced charges to the ground state before the next laser pulse hits the sample. The mobility is derived by the maximum signal height as the following equation[56],

$$\varphi\Sigma\mu = \frac{\Delta G_{max}}{\beta e I_0 F_A}$$

where $I_0$ is the number of photons per unit area per pulse, β a geometric factor related to the microwave cell, e is the elementary charge and $F_A$ the fraction of the light absorbed by the sample at the excitation wavelength used. We assume φ to be unitary for the low exciton binding energy of the material.

Time-correlated single photon counting (TCSPC) measurements were performed with a home-built setup equipped with PicoQuant PDL 828 ''Sepia II'' and a PicoQuant HydraHarp 400 multichannel picosecond event timer and TCSPC module. A 640 nm pulsed laser (PicoQuant LDH-D-C-640) with a repetition rate of 2 MHz was used to excite the sample. A Thorlabs FEL-700 long-pass filter was used to remove the excitation laser.

**Supporting Information**

SEM at lower magnification of AS sample; polarized optical microscopy images of FIRA with polarizer and analyzer at different angles; EBSD measurement of a FIRA cluster; crystal orientation distribution along x-,y- direction for AS and FIRA sample; IQ and IPF overlay showing orientation along x- and y-direction for the AS sample; grain size distribution for AS and FIRA obtained from EBSD measurement; spatially resolved PL of bigger area of FIRA sample; AFM of different cluster boundary region in the FIRA sample; Lifetime map of FIRA sample at the cluster boundary region; Normalized PL spectra of five random regions in the AS and FIRA sample; EBSD geometry, Hough transformation and pattern indexing; Additional thickness travelled by light as function of emission wavelength in $MAPbI_3$ fabricated by FIRA;

**Conflict of Interest**

The authors declare no competing interests.

**Acknowledgements**

The authors thank Erik C. Garnett for providing feedback on the paper and Andries Lof for the technical support during the EBSD measurements. This work is part of the research program of the Netherlands Organization for Scientific Research (NWO). S.S. and M.S. thank the Adolphe Merkle Institute at the University of Fribourg, Switzerland, for past support.

**TOC**

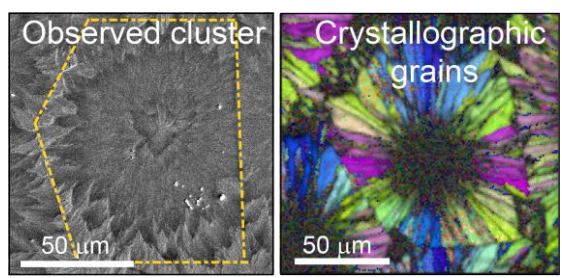